\documentclass[12pt]{article}

\usepackage{scicite}

\usepackage{times}
\usepackage[pdftex]{graphicx}
\usepackage[figuresright]{rotating}%
\usepackage{amsmath,amssymb,amsfonts}
\usepackage{textcomp}
\usepackage{xcolor}
\usepackage{braket}
\usepackage{multirow}

\def\citepunct{,}
\def\citedash{--}

\def\BibTeX{{\rm B\kern-.05em{\sc i\kern-.025em b}\kern-.08em
    T\kern-.1667em\lower.7ex\hbox{E}\kern-.125emX}}

\newcommand{\system}{QiankunNet} 

\newcommand{\boldx}{\bold{x}}

\newcommand{\im}{{\rm i}}

\newcommand{\shang}[1]{#1} 

\newenvironment{sciabstract}{%
\begin{quote} \bf}
{\end{quote}}

\title{Solving Schr\"odinger Equation with a Language Model}

\author
{Honghui Shang,$^{1\dagger}$ Chu Guo,$^{2\dagger}$ Yangjun Wu,$^{1}$ Zhenyu Li,$^{1}$ Jinlong Yang$^{1\ast}$\\
\\
\normalsize{$^{1}$Key Laboratory of Precision and Intelligent Chemistry,}\\
\normalsize{University of Science and Technology of China, Hefei, China}\\
\normalsize{$^{2}$Hunan Normal University, Changsha, China}\\
\\
\normalsize{$^\dagger$These authors contribute equally to this work.}\\
\normalsize{$^\ast$To whom correspondence should be addressed; E-mail:  jlyang@ustc.edu.cn.}\\
}

\date{}

\begin{document} 


\baselineskip24pt


\maketitle


\begin{sciabstract}
\shang{The fundamental many-electron Schr\"odinger equation is solved straightforwardly with QiankunNet, a neural network quantum state~(NNQS) framework based on generative Transformer architecture along with a batched autoregressive sampling method tailored for this Transformer-based ansatz in quantum chemistry calculations. This approach significantly improves the accuracy and efficiency of first-principles calculations compared to previous fermionic ansatz methods. The intricate quantum correlations are effectively captured by incorporating an attention mechanism into the methodology. Additionally, a batched autoregressive sampling strategy is employed to substantially enhance the sampling accuracy and efficiency. Furthermore, QiankunNet can incorporate a pre-training stage, where the truncated configuration interaction solution is embedded into the variational ansatz, ensuring high expressiveness and further boosting computational efficiency. QiankunNet showcases the power of the Transformer-based language model in achieving unprecedented efficiency in quantum chemistry calculations, opening up new avenues for chemical discovery and demonstrating the potential to solve the large-scale Schr\"odinger equation with modest computational cost.		
}
\end{sciabstract}



\section*{Introduction}
Recently, the Transformer architecture has revolutionized the field of natural language processing~(NLP), giving rise to large language models (LLMs) with unprecedented  capabilities~\cite{Transformer2017,radford2018gpt,radford2019gpt2,Brown2020}. Intrinsic versatility and adaptability of the Transformer architecture serve as the backbone of these LLMs, making it a powerful tool for addressing complex challenges, and its application are extending far beyond language processing, permeating various domains and revolutionizing numerous fields. Particularly, deploying this architecture within the scientific field harbors potential to propel us towards untapped frontiers of innovation and discovery.  While the Transformer has demonstrated exceptional performance in tasks like image recognition\cite{dosovitskiy2020image,BEiT2021}, protein representation and protein design\cite{Jumper2021-alphafold2,Unsal2022,Vu2023}, global weather 
forecasting\cite{Bi2023}, its potential benefits in the realm of quantum physics --- the language of our nature at the microscale --- remain largely unexplored. Therefore, it is both fundamentally intriguing and practically important to explore the ability of Transformer-based language models to solve the  Schr\"odinger equation, which is expected to unlock new levels of understanding and advancements in quantum physics.

In principle, the electronic structure and properties of all materials can be determined by solving the Schr\"odinger equation to obtain the wave function. However, in practice, it is a big challenge to find a general approach to reduce the exponential complexity of the many-body wave function and extract its essential features. Various methods have been developed to solve the Schr\"odinger equation for realistic systems. Considering the fermionic nature of electrons, Slater determinants are used to represent the electron systems, ensuring that the wave function obeys exchange antisymmetric symmetries. While the full configuration interaction (FCI) method provides a comprehensive approach to obtain the exact wavefunction, the exponential growth of the Hilbert space limits the size of feasible FCI simulations. To approximate the exact energy, several strategies have been devised, including perturbation theories \cite{Perturbation, MP2}, \shang{the trunacted configuration interaction which takes into account arbitrary linear combinations of excitations up to a certain order~\cite{MCSCF}, the coupled-cluster (CC) method which takes into account certain nonlinear combinations of excitations up to a certain order~(e.g. CCSD, CCSD(T))~\cite{Bartlett2007}, the density matrix renormalization group (DMRG) algorithm~\cite{White1992,White1993} which uses the one-dimensional matrix product state wave function ansatz, or the variational Monte Carlo (VMC) method~\cite{McMillan1965,FoulkesRajagopal2001,AustinLester2012} which works for any wave function ansatz.} However, these methods can fail in numerous cases, mostly due to the limited expressive power of the wave function ansatz. 

In 2017, Carleo and Troyer proposed their seminal work on the neural network quantum state (NNQS) algorithm, which introduced a groundbreaking approach for tackling many-spin systems within the exponentially large Hilbert space~\cite{CarleoTroyer2017}. \shang{The main idea behind NNQS is to parameterize the quantum wave function with a neural network and optimize its parameters stochastically using the variational Monte Carlo (VMC) algorithm.
	They also demonstrated the neural network ansatz, when dealing with many-body quantum states, has a bigger expressive power compared to the tensor network states~\cite{DengSarma2017,GlasserCirac2018,SharirCarleo2022,GaoDuan2017,Huang2021}, and its computational cost typically scales polynomially~\cite{HermannNoe2020}. Moreover, as a general feature of Monte Carlo methods, large-scale parallelization can provide notable performance advantages.} The NNQS method has been applied to both first quantized~\cite{PfauFoulkes2020,HermannNoe2020,vonglehn2022psiformer} and second quantized formalism~\cite{ChooCarleo2020,BarrettLvovsky2022,Zhao2023,PRL2023}. \shang{ The first quantization method has the advantage of already featuring a full basis set limit, while the discrete basis approach (second quantization method) has the advantage of easily enforcing boundary conditions and fermionic symmetry. In addition, the first quantization approach uses the Markov chain (MC) sampling method, as it samples the individual electrons and their coordinates in real space. However, such sampling can be inefficient in certain situations with a low acceptance rate and can result in correlated samples, making it very challenging for the first quantization approach to scale up to large systems.} \shang{
	In the second quantization formalism, the number of terms in the Hamiltonian scales as the fourth power of the number of spin orbitals, consequently, in large-scale applications, the evaluation of the local energy in VMC becomes very time-consuming, posing a major limitation to practical scalability~\cite{ChooCarleo2020}. Additionally, the neural network complexity of current state-of-the-art second quantization methods~\cite{BarrettLvovsky2022} scales with the number of spin orbitals, leading to a dramatic increase in computational demands and memory usage as the system size grows, further makes scalability a tremendous challenge. At the same time, molecular systems exhibit intricate chemical bonding patterns, necessitating  models with high expressibility to accurately capture the intricate electronic correlations~\cite{Zhao2023}, for practical applications, it is crucial to employ models that can be readily extended to deep architectures to improve their expressive power.}


\shang{In this work, we develop~\system(Qiankun means ``heaven and earth"), 
	which seamlessly integrates cutting-edge Transformer architectures and autoregressive sampling strategies. The Transformer model's self-attention mechanism enables it to effectively capture intricate electron correlations, and the number of Transformer network's  parameters is independent of the number of the spin orbitals. Moreover, QiankunNet's batched autoregressive sampling circumvents the issues associated with MCMC, facilitating efficient and scalable computations for large quantum systems.  A highly efficient local energy evaluation is performed in a parallelized manner over batches of samples, leveraging a highly compressed data structure for the Hamiltonian.} Furthermore, the neural network architecture can be pre-trained to encode the physics of valid wavefunctions such as the configuration interaction method. These techniques are essential to obtain a method that not only achieves high accuracy but also converges robustly with improved computational efficiency. Through several test systems, we demonstrate that QiankunNet substantially outperforms available wavefunction ansatzes. Our method open a new avenue for accurate large scale electronic structure calculations.

\section{QiankunNet: A Transformer-based Ansatz for Efficient Quantum Chemistry Calculations}\label{sec2}
In the second quantized formalism, with a basis set~(single-electron quantum states or spin-orbitals) introduced, the many-electron wave function can be written as a linear combination of configurations
\begin{equation}
	|\Psi\rangle =\sum_{\boldx}   \langle \boldx |\Psi \rangle |\boldx \rangle =  \sum_{\boldx} \Psi(\boldx)  |\boldx \rangle
\end{equation}
where each configuration is represented by an occupation number vector (`configuration string')  $|\boldx \rangle=\{x_1,x_2,...x_N\}$ with $x_i \in \{ 0, 1\}$ denoting whether the i-th spin orbital is occupied or not. 
Then the second quantized electronic Hamiltonian can be mapped to a linear combination of Pauli strings \cite{Jordan1928}.

In the context of NLP, a configuration string can be considered as a sentence with each word is either 0 or 1. Then, the challege in sloving the Schrodinger equation, the Hilbert space becomes exponentially large with the size of the system ($2^N$ for $N$ spin orbitals), is exactly what we encounter in NLP where the number of possible sentences is also exponentially large. By leveraging state-of-the-art language modeling techniques employing Transformer-based architectures, such as ChatGPT~\cite{radford2018gpt,radford2019gpt2}, sentences can be predicted with remarkable precision. Compared to other deep neural networks like convolutional neural networks (CNNs), Transformers have a remarkable ability to capture long-range dependencies while maintaining a constant number of operations~\cite{Transformer2017}. Drawing inspiration from the remarkable achievements of Transformer-based architectures in NLP and various machine learning tasks, we adopt a customized deep neural network based on  Transformer  as the wave function ansatz.

\begin{sidewaysfigure*}[p]
	\centerline{\includegraphics[width=1\columnwidth]{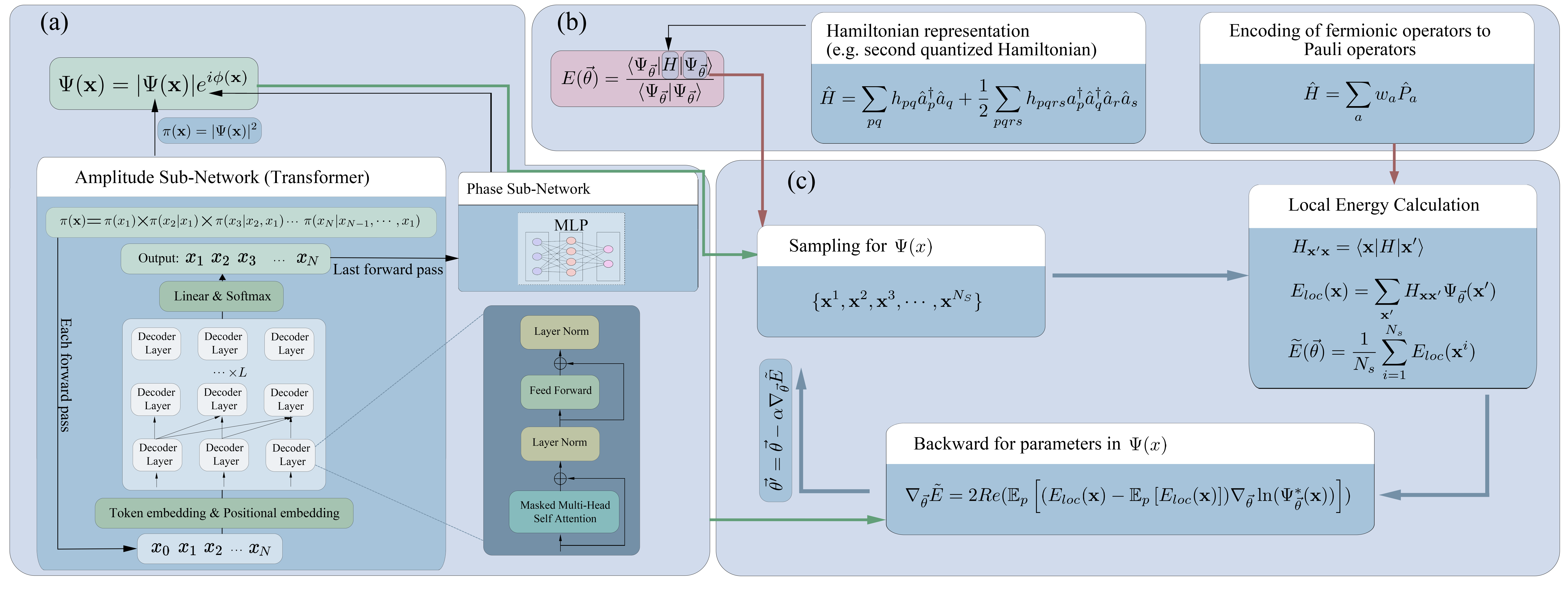}}
	\setlength{\abovecaptionskip}{0.05cm}
	\caption{\textbf{QiankunNet: the NNQS-Transformer pipeline.} (a) The GPT-style decoder-only Transformer architecture  for the electron wave function ansatz, where the Transformer is used for the amplitude and multi-layer perceptron is used for the phase. (b) Pre-processing: The Hamiltonian of the molecular systems can be expressed as quantum one-electron states. Then, the Hamiltonian is mapped to spin operators using Fermion-to-qubit encoding. (c) Schematic flowchart of the NNQS calculation.}
	\label{fig:nnqs}
\end{sidewaysfigure*}

QiankunNet consists of two sub-networks: the amplitude sub-network and the phase sub-network.
\begin{align}
	\Psi(\boldx) = |\Psi(\boldx)| e^{\im \phi(\boldx)},
\end{align}
The amplitude sub-network is constituted by a Transformer decoder to represent the probability $|\Psi(\boldx)|^2$, while the phase sub-network $\phi(\boldx)$ is composed of a multi-layer perceptron (MLP), as shown in Fig.~\ref{fig:nnqs}(a). 
The input to our network is a token/qubit vector $\boldx=\{x_1,x_2,...,x_i\}$ of length $i$, and the amplitude sub-network produces the probability distribution of $\boldx$. 
The mathematical formulation of QiankunNet is as follows:
\begin{equation}
	\begin{aligned}
		h_0 = XW_e+W_p  \\
		h_j = \mathrm{decoder\_layer}(h_{j-1}), \quad j\in [1,n]  \\
		|\Psi(\boldx)|^2 \sim \mathrm{softmax(h_n,W_{head})} \\
		\rm{Phase(\boldx) = MLP(\boldx)}
	\end{aligned}
\end{equation}
Here, $W_e$, $W_p$, and $W_{\rm head}$ denote the token, position, and output embedding matrices, respectively. The initial state of the decoder $h_0$ is calculated using the input token vector and the position embedding matrix. Each $h_j$ represents the j-th hidden state within the decoder layers. We have only used decoders since our problem is unsupervised/self-supervised. Finally, the amplitude of the quantum state $|\Psi(\boldx)|$ is derived by applying a softmax activation function to the output of the final decoder layer and the output embedding matrix. The phase of the quantum state, denoted as $\rm{Phase(\boldx)}$, is computed by feeding the input token vector into the MLP.   

VMC is used to optimize QiankunNet using energy as the loss function. To estimate the energy, a set of configuration strings are sampled  according to the underlying distribution $|\Psi(\boldx)|^2$. One common approach for such a sampling is Markov chain Monte Carlo (MCMC). However, its dependency on a large volume of samples, coupled with an extremely low acceptance probability, inevitably leads to the inefficiency of sampling, thereby posing a significant bottleneck when it comes to scaling up to larger systems.
For example, in the case of applying restricted Boltzmann machines (RBMs) to quantum chemistry, Choo et al.~\cite{ChooCarleo2020} observed that the number of samples generated from the MCMC method was limited to 10$^6$ due to the heavy computational cost, while the acceptance ratios were as low as 0.1\%.

\begin{figure*}
	\centerline{\includegraphics[width=1.0\columnwidth]{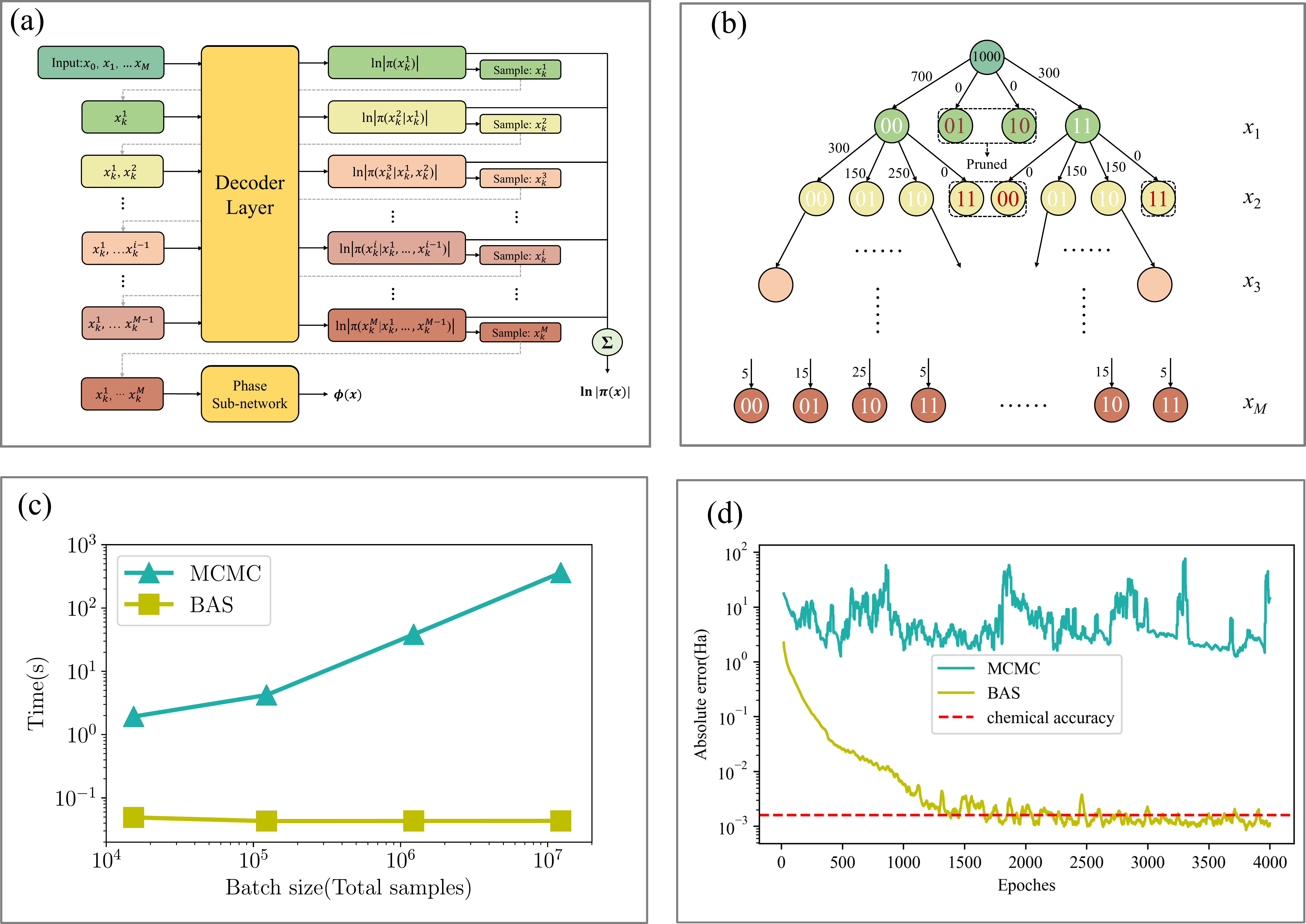}}
	\caption{\textbf{Comparison between different sampling algorithms. }(a) The autoregressive sampling~(AS) algorithm which generate one sample per run. (b) The batched autoregressive sampling~(BAS) algorithm which generates $N_s$ samples per run. Each circle in (b) corresponds to a particular local sampling outcome, and the number on the edge pointing to the circle means the weight. $N_s$ can be chosen to be any number ($N_s=1000$ is used as an example). (c) The time comparison of H$_2$O system for Markov Chain Monte Carlo~(MCMC) and BAS sampling methods. In both cases, the \system~is used for the wave function ansatz. \shang{(d) The convergence comparison of H$_2$O between MCMC and BAS sampling methods. $N_s=10^{5}$ is used as an example. }}
	\label{fig:sampling}
\end{figure*}

Notice that the amplitude part of QiankunNet, which determines the probability distribution, is an autoregressive network. Therefore, as illustrated in Fig.~\ref{fig:sampling}(a), the autoregressive sampling~(AS) method can be straightforwardly used to produce a single sample per run with a 100\% acceptance ratio\cite{Sharir2020}. In this study, we adopt a more efficient technique called batched autoregressive sampling (BAS) proposed in Ref.~\cite{BarrettLvovsky2022}. As shown in Fig.~\ref{fig:sampling}(b), the main idea of BAS is to generate a batch of samples instead of a single one at each sampling step. By sampling numbers of occurrences for each unique string instead of sampling configuration strings directly, the computational cost only scales with the number of unique samples(N$_u$) instead of the overall sample batch size. 

In Fig.~\ref{fig:sampling}(c), we showcase a comparison between the performance of MCMC sampling and the BAS method for H$_2$O system. From this comparison, it becomes evident that the computational time of the MCMC method increases as the number of samples grows, and the low acceptance probability associated with the Metropolis-Hastings algorithm renders it inefficient for larger sample sizes. In contrast, the computational time of the BAS method exhibits a nearly constant behavior, since it is primarily dependent on the number of unique samples. A computational efficiency improvement of several orders of magnitude is achieved compared to the MCMC method. \shang{Figure~\ref{fig:sampling}(d) illustrates that, with the same number of samples (e.g., 10$^5$), the BAS method achieves chemical accuracy within 1500 steps, whereas the MCMC method fails to attain chemical accuracy even after an extensive number of steps.}

With the BAS sampling method, we can optimize the \system~ ansatz in an unsupervised fashion using the variational principle for the total electronic energy. It is thus an \textit{ab initio} approach without requiring  pre-existing data. Nevertheless, achieving a well-performing representation of the wave function typically requires a significant number of iterations, given the large number of parameters present in neural models, which are challenging to quickly learn. Therefore, we can improve the performance of QiankunNet via supervised pre-training, using methods such as configuration interaction.

\section*{Numerical results}
\begin{figure}
	\centerline{\includegraphics[width=\columnwidth]{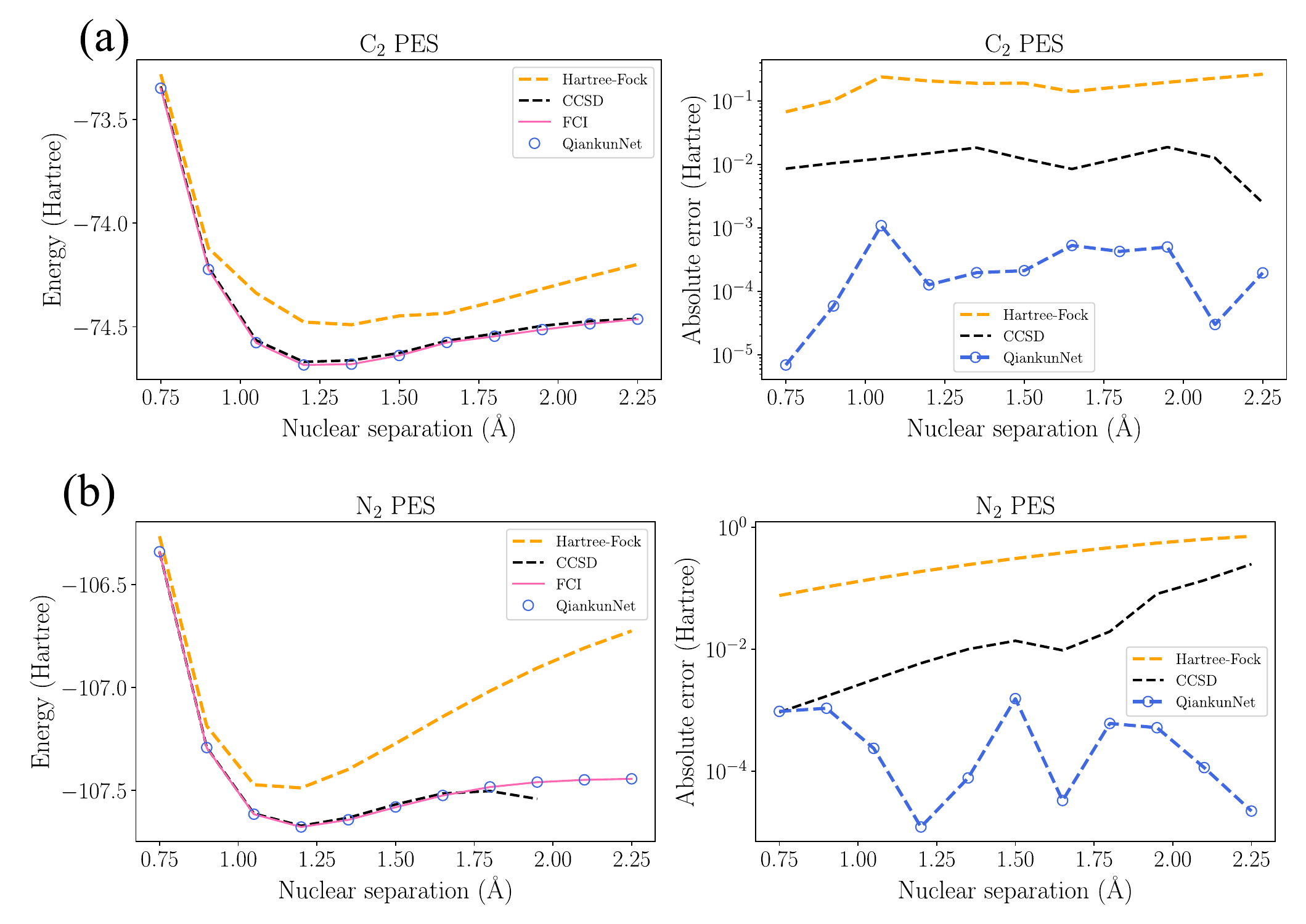}}
	\setlength{\abovecaptionskip}{0.05cm}
	\caption{\textbf{Potential energy curves for C$_2$ and N$_2$.} Comparison of the energies obtained using QiankunNet and other traditional quantum chemistry approaches for (a) C$_2$ and (b) N$_2$, as a function of the nuclear separation. \system~outperforms all other approximation techniques. It agree with FCI results well even at structures where CCSD failed due to the presence of strong correlations.   }
	\label{fig:C2-and-N2}
\end{figure}

\begin{figure*}[t]	      
	\begin{center}
		\includegraphics[width=1.0\columnwidth]{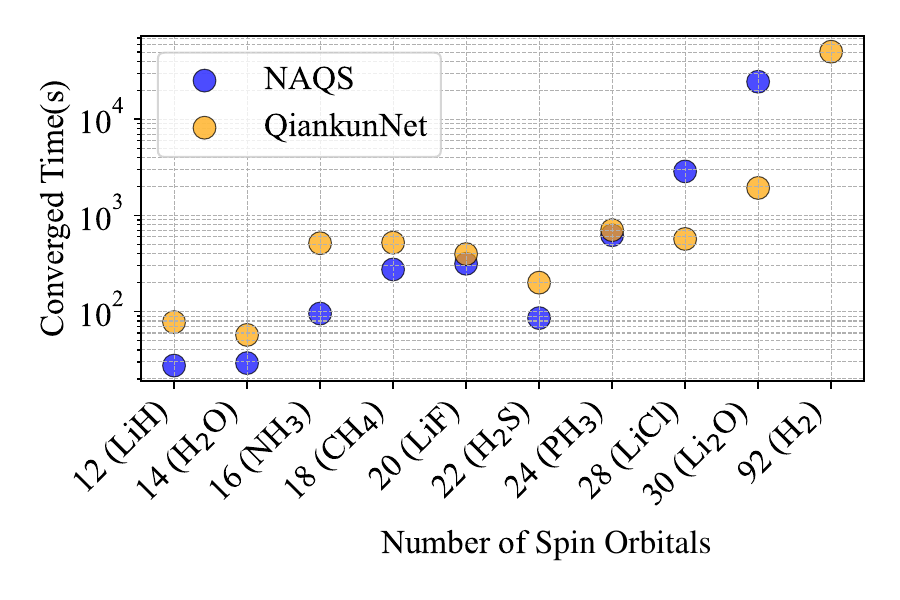}
	\end{center}
	\caption{The  performance comparison between NAQS and QiankunNet in simulating real chemical systems. The  measurements here refer to the time during which these two methods perform calculations until convergence is achieved.   }
	\label{fig:compare}
\end{figure*}

To demonstrate the performance of QiankunNet, we compute the ground state energies of several small molecules with the minimal Gaussian basis set and compare them to results of other methods, such as Hartree–Fock~(HF) and coupled cluster with up to double excitations (CCSD).  We also report results from NNQS methods using other neural-network ansatz~(NAQS and MADE). As listed in Table~\ref{tab:mol}, QiankunNet achieves the chemical accuracy compared to the ground-truth FCI result (within an absolute error of $1.6\times 10^{-3}$ Hartree compared to FCI) for 16 molecules and it exhibits consistently higher accuracy compared to other methods.QiankunNet can be successfully used to obtain the full potential energy surface, as demonstrated by C$_2$ and N$_2$ (Fig.~\ref{fig:C2-and-N2}). Even in cases where the CCSD baselines fail, QiankunNet gives very accurate results.  We systematically investigate the impact of the number of calculation parameters and the pre-training model, with the results presented in the Supplementary Materials.  
\shang{When comparing with other second quantized NNQS approaches, the Transformer-based neural network adopted in QiankunNet can exhibits heightened accuracy. For example, the second quantized approached such as MADE method cannot achieve the chemical accuracy for C$_2$ and N$_2$ systems, while QiankunNet can achieving an accuracy two orders of magnitude greater. Another advanced second quantized method, NAQS, employs a multilevel perceptron (MLP) augmented with hard-coded pre-and postprocessing steps to maintain the autoregressive property. However, the complexity of NAQS's neural network scales with the number of spin orbitals, leading to a dramatic increase in computational demands and memory usage as the system size grows. For instance, for a C$_2$H$_4$O molecule with 38 spin orbitals the GPU VRAM requirement soars to an unsustainable 454 GB, far surpassing the 80GB memory capacity of an A100 GPU. Conversely, QiankunNet incorporates a decoder-only Transformer, making the number of the network parameters independent of the number of the spin orbitals (our model can potentially be simultaneously used for systems with different number of spin orbitals), as a result, QiankunNet demonstrates a more favorable time scaling compared to NAQS, for small systems with fewer than 24 spin orbitals, NAQS calculation is faster, however, when assessing a system with 28/30 spin orbitals, like LiCl/Li$_2$O, QiankunNet boasts a computational speed 5/10 times faster than NAQS. And can extended to chemical systems up to 92 spin-orbitals, as illustrated in Fig.~\ref{fig:compare}. Beyond speed, our Transformer-based neural network also exhibits higher accuracy. }
For instance, in the Li$_2$O case, while NAQS captures 98.1\% of the electron correlation energy, QiankunNet recovers an impressive 99.5\% of the electron correlation energy.  We also perform the calculation for the C$_2$H$_4$O molecule comprising 38 spin orbitals, and our method yield an energy better than MADE approach~\cite{Zhao2023}.

\section*{Discussions} 

In this study, we have proposed \system~ by leveraging the transformer architecture into solving the many-electron Schr\"odinger equation, marking the first instance of a GPT-style language model successfully capturing the quantum electronic structure, which is the language of nature at the microscale. Following the success of applying transformer-based language models in human language prediction and protein sequence structure understanding, we have demonstrated that they can also lead to a deep comprehension of the patterns inherent in electron wave functions.
With their ability to process input sequences in parallel and efficiently handle long-range dependencies, transformer-based language models have surpassed previous architectures in terms of both training efficiency and accuracy.  With the highly effecient BAS technique, unsupervised training of QiankunNet via minimizing the total energy can be converged fast. If enough computational resources are provided (for example the amount used for training ChatGPT), some long standing challenges in chemistry, such as understanding the accurate electronic structure of the strongly correlated system iron molybdenum cofactor (FeMoco), are expected to be solved straightforwardly (Table S2).

We notice that there is a recent work by Glehn et al.~\cite{vonglehn2022psiformer} reporting a network called Psiformer which also use the attention mechanism to characterize interactions between electrons. However, Psiformer treats the first quantized Hamiltonian and encodes wavefunctions in continuous real space, which is less efficient compared to the discrete basis sets used in QiankuNet. At the same time, Psiformer employs an encoder-style neural network that lacks autoregressive properties, which makes it relies solely on the MCMC sampling method thus hard to be scaled up to large systems. For an in-depth comprehensive comparison between QiankunNet, Psiformer, and other state-of-the-art methods, please refer to the supplementary material and Table S1 within it.

This study provides a profound link between language modeling and the knowledge about electron behavior, further emphasizing the versatility and potential of these models in deciphering complex scientific phenomena.  We hold the belief that \textit{ab initio} approaches based on neural network wavefunctions, which is highly accurate, scalable, and with high computational efficiency, shall emerge as an important component within the quantum chemistry, facilitating more efficient electronic-structure calculations of complex molecular systems.



\bibliography{refs}

\begin{thebibliography}{10}

\bibitem{Transformer2017}
A.~Vaswani, {\it et~al.\/}, Attention is all you need.
\newblock {\it Proceedings of the 31st International Conference on Neural
  Information Processing Systems\/}, NIPS'17 (Curran Associates Inc., Red Hook,
  NY, USA, 2017), p. 6000–6010.

\bibitem{radford2018gpt}
A.~Radford, K.~Narasimhan, T.~Salimans, I.~Sutskever, {\it et~al.\/}, Improving
  language understanding by generative pre-training.
\newblock {\it OpenAI blog\/}  (2018).

\bibitem{radford2019gpt2}
A.~Radford, {\it et~al.\/}, Language models are unsupervised multitask
  learners.
\newblock {\it OpenAI blog\/} {\bf 1}, 9 (2019).

\bibitem{Brown2020}
T.~B. Brown, {\it et~al.\/}, Language models are few-shot learners.
\newblock {\it Proceedings of the 34th International Conference on Neural
  Information Processing Systems\/}, NIPS'20 (Curran Associates Inc., Red Hook,
  NY, USA, 2020).

\bibitem{dosovitskiy2020image}
A.~Dosovitskiy, {\it et~al.\/}, An image is worth 16x16 words: Transformers for
  image recognition at scale.
\newblock {\it arXiv:2010.11929\/}  (2020).

\bibitem{BEiT2021}
H.~Bao, L.~Dong, S.~Piao, F.~Wei, Beit: Bert pre-training of image
  transformers.
\newblock {\it ICLR 2022\/} (2022).

\bibitem{Jumper2021-alphafold2}
J.~Jumper, {\it et~al.\/}, {Highly accurate protein structure prediction with
  AlphaFold}.
\newblock {\it Nature\/} {\bf 596}, 583 (2021).

\bibitem{Unsal2022}
S.~Unsal, {\it et~al.\/}, {Learning functional properties of proteins with
  language models}.
\newblock {\it Nat. Mach. Intell.\/} {\bf 4}, 227 (2022).

\bibitem{Vu2023}
M.~H. Vu, {\it et~al.\/}, {Linguistically inspired roadmap for building
  biologically reliable protein language models}.
\newblock {\it Nat. Mach. Intell.\/} {\bf 5}, 485 (2023).

\bibitem{Bi2023}
K.~Bi, {\it et~al.\/}, {Accurate medium-range global weather forecasting with
  3D neural networks}.
\newblock {\it Nature\/}  (2023).

\bibitem{Perturbation}
T.~Helgaker, P.~J{\o}rgensen, J.~Olsen, {\it {Perturbation Theory}\/} (John
  Wiley and Sons, Ltd, 2000), chap.~14, pp. 724--816.

\bibitem{MP2}
C.~M\o{}ller, M.~S. Plesset, Note on an approximation treatment for
  many-electron systems.
\newblock {\it Phys. Rev.\/} {\bf 46}, 618 (1934).

\bibitem{MCSCF}
R.~Shepard, {\it The Multiconfiguration Self-Consistent Field Method\/} (John
  Wiley \& Sons, Ltd, 1987), pp. 63--200.

\bibitem{Bartlett2007}
R.~J. Bartlett, M.~Musia\l{}, Coupled-cluster theory in quantum chemistry.
\newblock {\it Rev. Mod. Phys.\/} {\bf 79}, 291 (2007).

\bibitem{White1992}
S.~R. White, Density matrix formulation for quantum renormalization groups.
\newblock {\it Phys. Rev. Lett.\/} {\bf 69}, 2863 (1992).

\bibitem{White1993}
S.~R. White, Density-matrix algorithms for quantum renormalization groups.
\newblock {\it Phys. Rev. B\/} {\bf 48}, 10345 (1993).

\bibitem{McMillan1965}
W.~L. McMillan, Ground state of liquid ${\mathrm{he}}^{4}$.
\newblock {\it Phys. Rev.\/} {\bf 138}, A442 (1965).

\bibitem{FoulkesRajagopal2001}
W.~M.~C. Foulkes, L.~Mitas, R.~J. Needs, G.~Rajagopal, Quantum monte carlo
  simulations of solids.
\newblock {\it Rev. Mod. Phys.\/} {\bf 73}, 33 (2001).

\bibitem{AustinLester2012}
B.~M. Austin, D.~Y. Zubarev, W.~A.~J. Lester, Quantum monte carlo and related
  approaches.
\newblock {\it Chem. Rev\/} {\bf 112}, 263 (2012).

\bibitem{CarleoTroyer2017}
G.~Carleo, M.~Troyer, Solving the quantum many-body problem with artificial
  neural networks.
\newblock {\it Science\/} {\bf 355}, 602 (2017).

\bibitem{DengSarma2017}
D.-L. Deng, X.~Li, S.~Das~Sarma, Quantum entanglement in neural network states.
\newblock {\it Phys. Rev. X\/} {\bf 7}, 021021 (2017).

\bibitem{GlasserCirac2018}
I.~Glasser, N.~Pancotti, M.~August, I.~D. Rodriguez, J.~I. Cirac,
  Neural-network quantum states, string-bond states, and chiral topological
  states.
\newblock {\it Phys. Rev. X\/} {\bf 8}, 011006 (2018).

\bibitem{SharirCarleo2022}
O.~Sharir, A.~Shashua, G.~Carleo, Neural tensor contractions and the expressive
  power of deep neural quantum states.
\newblock {\it Phys. Rev. B\/} {\bf 106}, 205136 (2022).

\bibitem{GaoDuan2017}
X.~Gao, L.-M. Duan, Efficient representation of quantum many-body states with
  deep neural networks.
\newblock {\it Nat. Commun.\/} {\bf 8}, 662 (2017).

\bibitem{Huang2021}
Y.~Huang, J.~E. Moore, Neural network representation of tensor network and
  chiral states.
\newblock {\it Phys. Rev. Lett.\/} {\bf 127}, 170601 (2021).

\bibitem{HermannNoe2020}
J.~Hermann, Z.~Sch{\"a}tzle, F.~No{\'e}, Deep-neural-network solution of the
  electronic schr{\"o}dinger equation.
\newblock {\it Nat. Chem.\/} {\bf 12}, 891 (2020).

\bibitem{PfauFoulkes2020}
D.~Pfau, J.~S. Spencer, A.~G. D.~G. Matthews, W.~M.~C. Foulkes, Ab initio
  solution of the many-electron schr\"odinger equation with deep neural
  networks.
\newblock {\it Phys. Rev. Res.\/} {\bf 2}, 033429 (2020).

\bibitem{vonglehn2022psiformer}
I.~von Glehn, J.~S. Spencer, D.~Pfau, A self-attention ansatz for ab-initio
  quantum chemistry.
\newblock {\it arXiv:2211.13672\/}  (2022).

\bibitem{ChooCarleo2020}
K.~Choo, A.~Mezzacapo, G.~Carleo, Fermionic neural-network states for ab-initio
  electronic structure.
\newblock {\it Nat. Commun.\/} {\bf 11}, 2368 (2020).

\bibitem{BarrettLvovsky2022}
T.~D. Barrett, A.~Malyshev, A.~Lvovsky, Autoregressive neural-network
  wavefunctions for ab initio quantum chemistry.
\newblock {\it Nat. Mach. Intelle.\/} {\bf 4}, 351 (2022).

\bibitem{Zhao2023}
T.~Zhao, J.~Stokes, S.~Veerapaneni, {Scalable neural quantum states
  architecture for quantum chemistry}.
\newblock {\it Machine Learning: Science and Technology\/}  (2023).

\bibitem{PRL2023}
L.~L. Viteritti, R.~Rende, F.~Becca, Transformer variational wave functions for
  frustrated quantum spin systems.
\newblock {\it Phys. Rev. Lett.\/} {\bf 130}, 236401 (2023).

\bibitem{Jordan1928}
P.~Jordan, E.~Wigner, {{\"{U}}ber das Paulische {\"{A}}quivalenzverbot}.
\newblock {\it Zeitschrift f{\"{u}}r Physik\/} {\bf 47}, 631 (1928).

\bibitem{Sharir2020}
O.~Sharir, Y.~Levine, N.~Wies, G.~Carleo, A.~Shashua, Deep autoregressive
  models for the efficient variational simulation of many-body quantum systems.
\newblock {\it Phys. Rev. Lett.\/} {\bf 124}, 020503 (2020).

\end{thebibliography}

\bibliographystyle{Science}

\section*{Acknowledgments}
\textbf{Funding:} This work is supported by the Strategic Priority Research Program of the Chinese Academy of Sciences,Grant No. XDB0450101, and by National Natural Science Foundation of China~(Grant No. T2222026, 22003073, 11805279, and 21825302). This work was supported by the Supercomputing Center of the USTC.
\textbf{Author contributions:}The project was conceived by J.Y. The manuscript was written by H.S., C.G., Z.L., and J.Y. The numerical simulations and analysis were performed by H.S. and Y.W. All authors discussed the results. H.S. and C.G. contributed equally to this work and are considered as co-first authors. 
\textbf{Data and materials availability:} The data that support the findings of this study are available from the corresponding author upon reasonable request.

\section*{Supplementary Materials}
Materials and Methods \\
Supplementary Text  \\
Figure S1 ; Tables S1 to S2  \\
References (1-10)



\clearpage


\begin{table*}
	\caption{\textbf{Ground state energies (in Hartree) calculated using QiankunNet with the STO-3G basis set.} The conventional Hartree-Fock (HF) and Coupled Cluster With Single And Double Excitation (CCSD), FCI results along with results from existing neural network quantum state (NNQS) methods such as NAQS \cite{BarrettLvovsky2022} and MADE \cite{Zhao2023}, are also listed for the purpose of comparison. N$_o$ denotes the number of spin orbitals, N$_e$ represents the total number of electrons, N$_f$ indicates the number of physically valid determinants with conserved N$_e$ and total spin S, N$_u$ indicates the number of unique samples and N$_h$ signifies the total number of Pauli strings in the Hamiltonian.} 
\centering
\scalebox{0.7}
{
	\begin{tabular}{llllllllllll}
		\hline
		\hline
		Molecule & N$_o$ & N$_e$ & N$_f$ & N$_u$ & N$_h$ & HF & FCI & CCSD & NAQS & MADE & \system \\ \hline
		F$_2$ & 20 & 18 & 1.00E+02 &  65  & 2951 & -195.6380  & -195.6611  & -195.6611  & -195.6611  & -195.6611  & -195.6611  \\ 
		HCl & 20 & 18 & 1.00E+02 &  53  & 5851 & -455.1360  & -455.1562  & -455.1562  & -455.1562  & -455.1562  & -455.1562  \\ 
		LiH & 12 & 4 & 2.25E+02 &  118  &  631 & -7.7674  & -7.7845  & -7.7845  & -7.7845  & -7.7845  & -7.7845  \\ 
		H$_2$O & 14 & 10 & 4.41E+02 & 244  & 1390 & -74.9644  & -75.0155  & -75.0154  & -75.0155  & -75.0155  & -75.0155  \\ 
		CH2 & 14 & 8 & 7.35E+02 & 345 &  2058 & -37.3754  & -37.5044  & -37.4157  & -37.5044  & -37.5044  & -37.5044  \\ 
		O$_2$ & 20 & 16 & 1.20E+03 & 304 &  2879 & -147.5513  & -147.7502  & -147.7027  & -147.7500  & -147.7500  & -147.7501  \\ 
		BeH$_2$ & 14 & 6 & 1.23E+03 & 577  & 2074 & -14.4432  & -14.4729  & -14.4727  & -14.4729  & -14.4729  & -14.4729  \\ 
		H$_2$S & 22 & 18 & 3.03E+03 &  809 & 9558 & -394.3114  & -394.3546  & -394.3546  & -394.3546  & -394.3546  & -394.3546  \\ 
		NH$_3$ & 16 & 10 & 3.14E+03 & 2322  & 4929 & -55.4548  & -55.5211  & -55.5209  & -55.5211  & -55.5210  & -55.5211  \\ 
		N$_2$ & 20 & 14 & 1.44E+04 & 2141  & 2239 & -107.4990  & -107.6602  & -107.6561  & -107.6595  & -107.6568  & -107.6602  \\ 
		CH$_4$ & 18 & 10 & 1.59E+04 & 9951  & 8480 & -39.7266  & -39.8063  & -39.8060  & -39.8062  & -39.8062  & -39.8062  \\ 
		C$_2$ & 20 & 12 & 4.41E+04 & 5429  & 2239 & -74.4209  & -74.6908  & -74.6745  & -74.6899  & \---  & -74.6904  \\ 
		LiF & 20 & 12 & 4.41E+04 & 7565  &  5849 & -105.1137  & -105.1662  & -105.1592  & -105.1662  & -105.1653  & -105.1661  \\ 
		PH$_3$ & 24 & 18 & 4.84E+04 & 9394 & 24369 & -338.6341  & -338.6984  & -338.6982  & -338.6984  & -338.6982  & -338.6983  \\ 
		LiCl & 28 & 20 & 1.00E+06 & 11634  &  24255 & -460.8273  & -460.8496  & -460.8476  & -460.8496  & -460.8481  & -460.8495  \\ 
		Li$_2$O & 30 & 14 & 4.14E+07 &  46755 & 20558 & -87.7956  & -87.8927  & -87.8855  & -87.8909  & -87.8856  & -87.8922 \\ 
		C$_2$H$_4$O & 38 & 24 & 2.54E+09 & 89095  & 137218 & -150.9276  & -  & -151.12047  & -  & -151.12048  & -151.12153 \\ 
		\hline \hline
	\end{tabular}
}
\label{tab:mol}
\end{table*}

\end{document}